\documentclass[reprint, longbibliography, superscriptaddress]{revtex4-1}
 \usepackage[colorlinks,citecolor=blue,linkcolor=blue]{hyperref}
 \usepackage{commath}

\newcommand{\be}{\begin{equation}}

\newcommand{\ee}{\end{equation}}
\newcommand{\bea}{\begin{eqnarray}}
\newcommand{\eea}{\end{eqnarray}}

\newcommand{\ba}{\begin{array}}
	\newcommand{\ea}{\end{array}}

\newcommand{\bl}{\begin{flalign}}
\newcommand{\enl}{\end{flalign}}

\newcommand{\half}{\frac{1}{2}}

\newcommand{\coh}[2]{\ket{#1}\bra{#2}}
\usepackage{siunitx}

\usepackage{graphicx}
\usepackage{amssymb}
\usepackage{epstopdf}
\usepackage{amsmath,dsfont}
\usepackage{bm}
\usepackage{braket}
\usepackage{hyperref, xcolor}
\usepackage{units}

\usepackage[final]{changes}

\newcommand{\eq}[1]{Eq.~\eqref{#1}}
\newcommand{\Eq}[1]{Equation~\eqref{#1}}
\newcommand{\fig}[1]{Fig.~\ref{#1}}
\newcommand{\Fig}[1]{Figure~\ref{#1}}

\newcommand{\around}{$\sim$}

\newcommand*{\rom}[1]{\expandafter\@slowromancap\romannumeral #1@}
\renewcommand{\bf}{\mathbf}
\newcommand{\mc}{\mathcal}

\newcommand{\proj}[1]{\ket{#1}\bra{#1}}
\newcommand{\bs}{\begin{split}}
	\newcommand{\es}{\end{split}}

\makeatletter
\newsavebox{\@brx}
\newcommand{\llangle}[1][]{\savebox{\@brx}{\(\m@th{#1\langle}\)}%
	\mathopen{\copy\@brx\kern-0.5\wd\@brx\usebox{\@brx}}}
\newcommand{\rrangle}[1][]{\savebox{\@brx}{\(\m@th{#1\rangle}\)}%
	\mathclose{\copy\@brx\kern-0.5\wd\@brx\usebox{\@brx}}}
\makeatother

\newcommand{\mol}{{(n)}}

\newcommand{\eV}[1]{\SI{#1}{\electronvolt}}

\begin{document}
\title{Manipulating Core-Excitations in Molecules by X-ray Cavities}
\author{Bing Gu}
\thanks{{These authors contributed equally to this work.}}
\affiliation{Department of Chemistry and Department of Physics and Astronomy, University of California, Irvine, 92697, USA}
\author{Artur Nenov}
\thanks{{These authors contributed equally to this work.}}
\affiliation{Dipartimento di Chimica Industriale "Toso Montanari", Universit\`a degli studi di Bologna, Viale del Risorgimento 4, 40136 Bologna, Italy.}
\author{Francesco Segatta}
\affiliation{Dipartimento di Chimica Industriale "Toso Montanari", Universit\`a degli studi di Bologna, Viale del Risorgimento 4, 40136 Bologna, Italy.}
\author{Marco Garavelli}
\email{marco.garavelli@unibo.it}
\affiliation{Dipartimento di Chimica Industriale "Toso Montanari", Universit\`a degli studi di Bologna, Viale del Risorgimento 4, 40136 Bologna, Italy.}
\author{Shaul Mukamel}
\email{smukamel@uci.edu}
\affiliation{Department of Chemistry and Department of Physics and Astronomy, University of California, Irvine, 92697, USA}

\begin{abstract}
Core-excitations on different atoms are highly localized and therefore decoupled. By placing molecules in an X-ray cavity the core-transitions become coupled via the exchange of cavity photons and form  {delocalized} hybrid light-matter {excitations known as} core-polaritons.  We {demonstrate these effects for the two inequivalent  carbon atoms in} 1,1-difluoroethylene.  Polariton signatures in the  X-ray absorption, two-photon absorption, and multidimensional four-wave mixing,  signals are predicted.
\end{abstract}

\maketitle


Hybrid light-matter states between the material polarization and cavity photon modes, known as polaritons, {are created} when the light-matter coupling strength $g$ is larger than the decay rate of the cavity mode and the decoherence rate of the molecular transition (the strong coupling regime).
 When the cavity mode is in the vacuum state, the effective coupling strength for {an assembly of} $N$ identical molecules is {$\kappa = g\sqrt{N}  \mu$},$g \equiv  \sqrt{\frac{\hbar \omega_\text{c}}{2\varepsilon_0 V_\text{c}}}$, where $\varepsilon_0$ is the electric permittivity of vacuum, $\omega_\text{c}$ the cavity frequency, {$\mu$ the transition dipole moment},  and $V_\text{c}$ is the cavity mode volume \cite{Purcell1946}. The {$\sqrt{N}$ factor is responsible for}  cooperative superradiance \cite{Dicke1954, Gu2020b}.
 {Cavity polaritons} {in the visible and infrared regime} have   long been studied in atoms and were recently {experimentally reported in} molecules \cite{Zhong2017, George2015, Shalabney2015a, Benz2016, Yang2018, Ebbesen2016, Zhong2016, Chikkaraddy2016}.   Molecular electronic and vibrational polaritons have been experimentally shown to alter the electronic, optical, and chemical properties of molecules  including photoisomerization,  electronic energy transfer, electron transfer, and ground-state reaction rates \cite{Ebbesen2016}. These findings had triggered intensive theoretical investigations \cite{Gu2020b, Gu2020a, Kowalewski2016, Bennett2016, Schafer2019, Galego2015, Shalabney2015a,Flick2017, Flick2018, Dorfman2018, Herrera2017a, Coles2014, FriskKockum2019, Vasa2018, Hertzog2019, Flick2017, Schwartz2013, Herrera2017, Martinez-Martinez2018, Sanvitto2016, Gu2020c, Mandal2019}.

Thin-film optical cavities  in the hard X-ray regime have been recently employed in the study of collective M\"{o}ssbauer  signals of $^{57}$Fe nuclei (\SI{14.4}{\kilo\electronvolt}) \cite{Rohlsberger2020, Rohlsberger2010, Rohlsberger2012} and tantalum L-edge X-ray spectra (\eV{9881})  \cite{Haber2019}. A  $ \sim \eV{41}$ effective light-molecule coupling strength for low-finesse X-ray cavities has been reported  \cite{Haber2019}.
X-ray cavities {in the soft X-ray regime} can be formed by alternating nanometer layers of materials with different indices of refraction,
{and are on the horizon.} {For high-reflectivity mirrors, the cavity photon modes satisfy $\del{n + \half}\lambda_n = L$, where $L$ is the cavity length and $\lambda_n$ is the wavelength of the cavity mode. For carbon K-edge (\around 300 eV), it corresponds to $L \sim \SI{10}{\nano\meter}$.}

Here we study molecular polariton effects in the X-ray regime whereby a high-finesse X-ray cavity mode couples to molecular core-excitations.
%
We demonstrate that localized excitations from inequivalent carbon core-orbitals in 1,1-difluoroethylene can be coupled by the exchange of an X-ray cavity photon, leading to hybrid core-excitation with X-ray cavity photon modes.
{ Rich exciton-polariton physics has been observed  in the optical regime. This includes long-range transport \cite{Rozenman2018, Chervy2016}, enhanced optical nonlinearity, modified chemical reactions, polariton lasers, optical transistors, and phase transitions \cite{Ebbesen2016}. Our study suggests that similar phenomena may be expected for core-polaritons in the X-ray regime. X-ray cavities  enable long-range transport of core-excitons or core-holes despite their highly localized nature as long as the light-matter coupling strength is stronger than their decay rates} \cite{Rozenman2018, Chervy2016}.
We predict signatures of core-polaritons  in the X-ray absorption spectrum,
in two-dimensional (2D) X-ray four-wave mixing signals: photon echo and double quantum coherence, {and in the two-photon absorption}. Time-domain 2D spectroscopic techniques provide a versatile tool for exploring the optical properties of matter \cite{Mukamel1995, Xiang2018}.
 Multidimensional X-ray spectroscopy enabled by X-ray lasers \cite{Pellegrini2016}  can \cite{Mukamel2013, Tanaka2002, Biggs2012} capture electron dynamics  on the attosecond (as) time scale,
and can reveal the correlations among core-excitations. 


We consider a system of $N$ molecules coupled to a single X-ray cavity mode described by the Hamiltonian  $H = H_\text{M} +  H_\text{CM} + H_\text{LM}(t) +  H_\text{C} $ where {the $n$-th molecular Hamiltonian} $H_\text{M}^\mol = \sum_{j \in \set{g, e, f}} \hbar \omega_{j} \proj{j^\mol}$,
{the cavity Hamiltonian} $H_\text{C}  =  \hbar \omega_\text{c} a^\dag a$, {and the cavity-molecule coupling}
$
 H_\text{CM}  = \sum_{n=1}^N - \bm\mu^{(n)} \cdot \hat{\bf E}\del{\bf r_n, t}
$.
{Here $\bm \mu^\mol$ is the transition dipole moment and $\hat{\bf E}(\bf r) = i\sqrt{\frac{\hbar \omega_\text{c}}{2\varepsilon_0V_\text{c}}} \bf e_\text{c} a e^{i\bf k_\text{c} \cdot \bf r } + \text{H.c.}$  is the electric field operator} where
 $a$ ($a^\dag$) is the boson annihilation (creation) operator for the cavity mode, $\bf k_\text{c}, \bf e_\text{c}$ are the cavity mode wave vector and polarization, respectively, H.c. stands for the Hermitian conjugate.
We focus on the  single- and double-core carbon K-edge excited states, labeled $e$ and $f$, respectively. Double-core excitations of the same carbon atom are  excluded{, as they are blue-shifted by tens of eV with respect to doubly core-excited states on different atoms \cite{Nenov2019}. This shift can be attributed to  the reduced {electron} shielding caused by the first core-excitation which shifts a second core-excitation {from the same atom} to the blue.}
The {electric-dipole} coupling $H_\text{LM}(t)$ describes the interaction of the molecules with external laser pulses.

For $N > 1$ and $\abs{\bf k_\text{c} \cdot (\bf r_n - \bf r_m)} \ll 1$, it is convenient to introduce the collective core-exciton states
$ \ket{E_{\alpha k}} = \frac{1}{\sqrt{N}} \sum_{n=1}^N  e^{i kn } \ket{g^{(1)} \cdots g^{(n-1)} e^\mol_\alpha g^{(n+1)}\cdots}
$
describing a superposition of a single excitation shared by all molecules and similarly $\ket{F_{\mu k}}$,
where $k = 2 \pi j/N, j = 0, \cdots, N-1$. {Here $\alpha$ and $\mu$ run over the singly and doubly excited states, respectively.}
{Up to double excitations,} the {cavity-molecule} coupling can be represented by
(see Sec. S1 for details)
\be
\begin{split}
H_\text{CM} &=  \sum_{\alpha}  \sqrt{N} \kappa_{e_\alpha g} \ket{E_{\alpha 0}}\bra{G} a + \sum_k \sum_{\mu, \alpha} \kappa_{f_\mu e_\alpha} \ket{F_{\mu k} } \bra{E_{\alpha k}}a  \\
+& \sum_{n \ne m} \sum_{\alpha, \gamma} \kappa_{e_\alpha g} \ket{e^\mol_\alpha e^{(m)}_\beta} \bra{e^{(m)}_\beta} a + \text{H.c.}
\label{eq:coupling}
\end{split}
\ee
where $\ket{G} = \ket{g^{(1)} \cdots g^{(N)}}$.
\Eq{eq:coupling} implies that the transition from the ground-state to the delocalized core-exciton states $\ket{E_{\alpha 0}}$ is enhanced by $\sqrt{N}$ whereas the coupling between excited states $ \ket{E_{\alpha k}}$ and $\ket{F_{\beta k}}$ does not show such cooperativity \cite{Gu2020b}.
 The bright state $\ket{E_{\alpha 0}}$ is invariant under  exchange of any two molecules. The  dark states ${\ket{E_{\alpha k}}, k \ne 0}$ do not contribute to the absorption spectrum. Nevertheless, the transitions between the single-polariton and the dark biexciton states are coupled to the cavity mode by the $\ket{F_{\beta k} } \bra{E_{\alpha k}}a + \text{H.c.}$ term even for $k \ne 0$.
 Note that {bright} polariton states can relax to the dark states due to e.g. vibronic couplings, disorder, and cavity loss \cite{Gu2020b, DelPo2020}.
 The third term in \eq{eq:coupling} represents the coupling between the singly and doubly core-excited states from different molecules (Sec. S1).

\begin{figure}[ht]
	\includegraphics[width=0.45\textwidth]{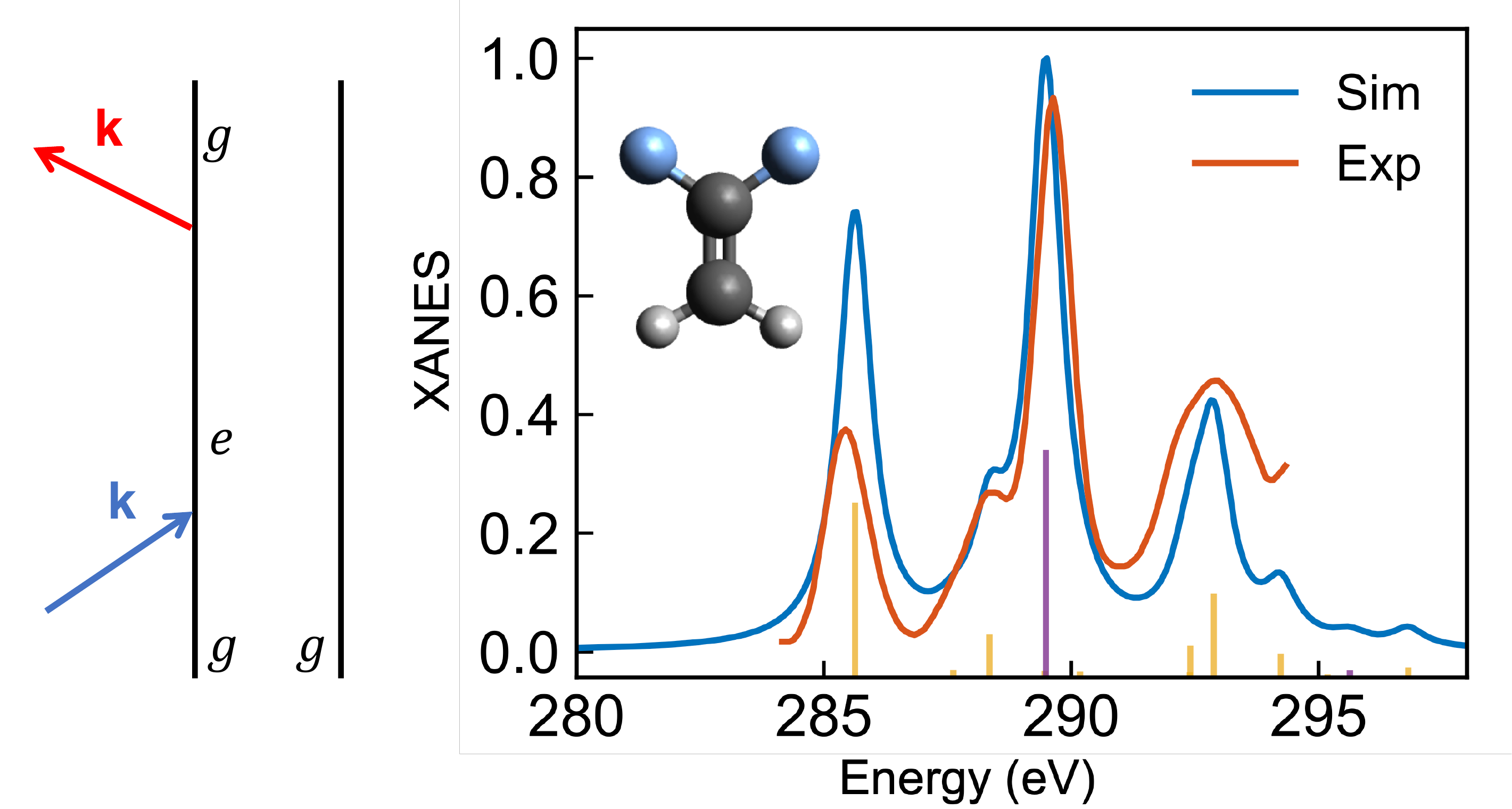}
	\caption{The XANES spectrum of 1,1-difluoroethylene and the corresponding (left) Feynman diagram. $\bf k$ denotes the incoming pulse wave vector. (right)  The transitions involving the carbon K-edge in CH\textsubscript{2} (CF\textsubscript{2})  are represented by yellow (purple) sticks. The agreement with experiment \cite{McLaren1987} are excellent.}
	\label{fig:bare}
\end{figure}

 \Fig{fig:bare} depicts the ground state structure of 1,1-difluoroethylene optimized at {the M\o{}ller-Plesset second-order perturbation level}, and compares the simulated and experimental X-ray absorption near edge structure (XANES) {in the [280, 296] eV spectral range}. This molecule has two inequivalent carbon atoms with bound pre-edge transitions {separated by a few eV}. The electronic structure computations are detailed in Sec. S2, and the  spectroscopic simulations in Sec. S3 \footnote{See Supplemental Material for details of the electronic structure and spectroscopic computations, which includes Refs. \cite{Forsberg1997, Malmqvist1990, Lundberg2019, Andersson1990, Sauri2011, Roca-Sanjuan2011, Ghigo2004, Zobel2017, Zobel2017, Pedersen2009, Kowalewski2017, Roos2004, Fdez.Galvan2019, Aquilante2020, Dorfman2012}.} The simulated XANES spectrum {(without any shift)} {is in excellent agreement} with experiment in the $\sbr{280, 296}$ eV spectral range.
The spectrum has four main features. The \SI{285.6}{\electronvolt} and \SI{289.5}{\electronvolt} {peaks} are associated with  excitations from the {$1s$ core orbitals of the} carbon atoms in the CH$_2$ and CF$_2$ groups to the anti-bonding $\pi^*$ orbital, respectively.  A broader peak at \SI{293}{\electronvolt}, arises from a pair of close lying transitions from CH\textsubscript{2} to Rydberg (Ry) orbitals. Finally, we find a red shoulder to  the \SI{289.5}{\electronvolt} band at \SI{288.4}{\electronvolt}, associated with a transition from CH\textsubscript{2} to a $\sigma^*$ anti-bonding orbital localized in the CH$_2$ fragment. The {$\sim$ \eV{4}} energy splitting  between the CH\textsubscript{2} $\rightarrow \pi^*$ and CF\textsubscript{2} $\rightarrow \pi^*$ peaks shows that  functionalization with electron withdrawing groups such as fluorine makes core-excitations more energy-costly, thus inducing a few eV blue-shift.  The K-edge spectrum is dominated by the core-excitations of CH\textsubscript{2}.


\begin{widetext}

\begin{figure}[ht]
\centering
	\includegraphics[width=0.6\textwidth]{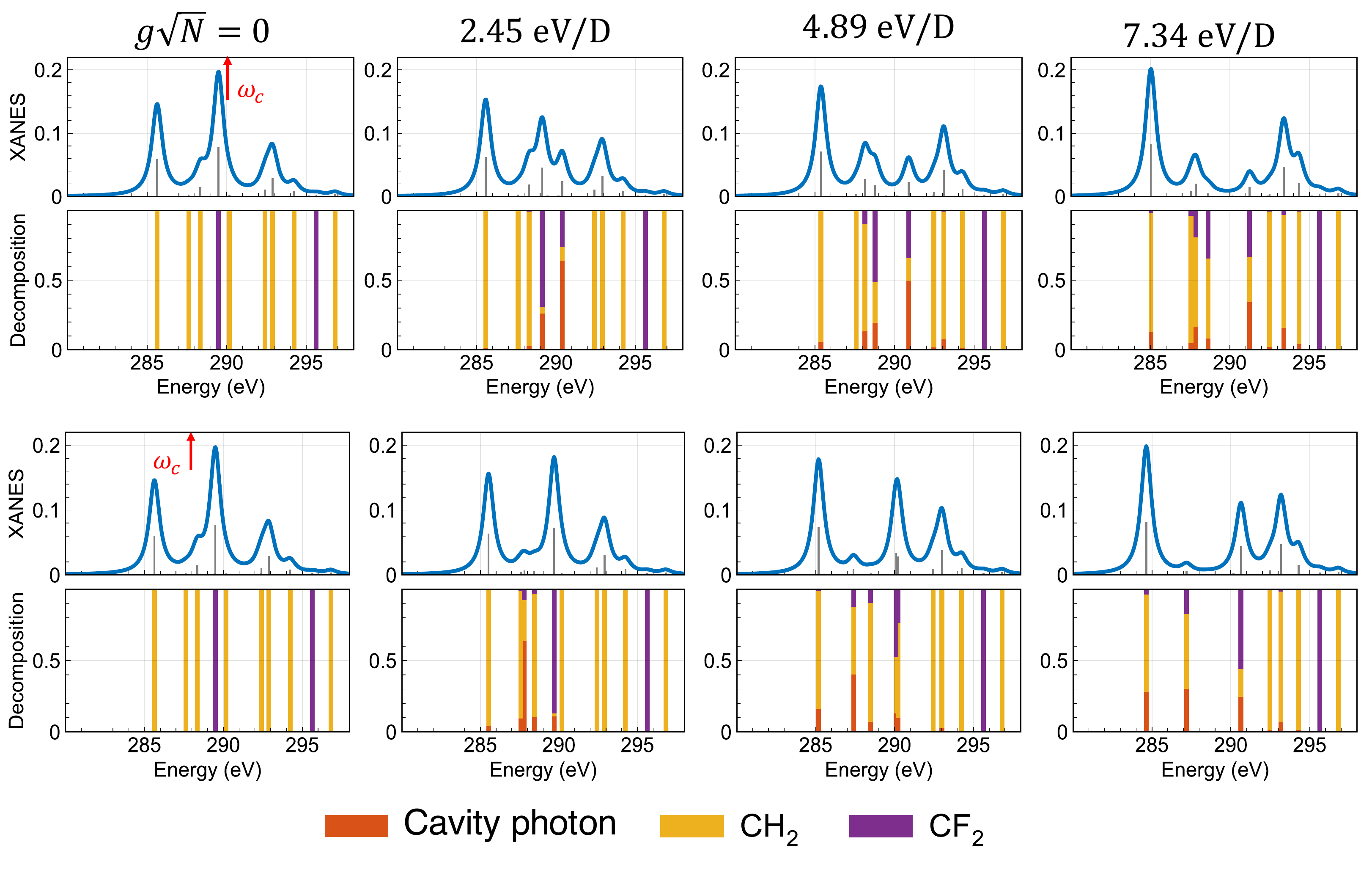}
	\caption{XANES of 1,1-difluoroethylene in an X-ray cavity for varying coupling strength. Lower panels show the decomposition of each polariton state into CH\textsubscript{2}, CF\textsubscript{2} and photon components. The top row is for cavity frequency $\omega_\text{c} = \eV{290}$ close to a specific transition, and the bottom row for cavity frequency $\omega_\text{c} = \eV{288}$ detuned from the main core-transitions. {The dependence on $N$ is solely through the  (collective) coupling strength $g\sqrt{N}$. }}
	\label{fig:xas}
\end{figure}

\end{widetext}

In the X-ray cavity, the core-excitations are modified by coupling to the cavity mode. \Fig{fig:xas} (top) {illustrates} the XANES of core-polaritons at cavity frequencies $\omega_\text{c} = \eV{290}$ close to the  CF\textsubscript{2} $\rightarrow \pi^*$ excitation for varying coupling strength. At {$g\sqrt{N} = 2.45$ eV/Debye} (eV/D), we observe a vacuum Rabi splitting of two polariton peaks. {The transition dipole is in the order of 0.1 D.} The Rabi splitting is increased with \deleted{$g$}{the coupling strength}, and  the lower polariton further mixes with CH\textsubscript{2} excitations leading to enhancement and redshift of the CH\textsubscript{2} $\rightarrow \pi^*$ transition.

To unveil the polaritonic nature of the core-excitations in the X-ray cavity, we have decomposed each polariton state into the CH\textsubscript{2}, CF\textsubscript{2}\ and the cavity photon components. These are depicted in the lower panels in \fig{fig:xas}. Since the core-excitations localized at CH\textsubscript{2} and CF\textsubscript{2} are decoupled, each excitation is either purely CH\textsubscript{2} or CF\textsubscript{2} type.  To decompose the polariton states,
we introduce the projection operators $\mc{P}_\sigma = \sum_{\alpha \in \sigma} \proj{e_\alpha}$ where $\sigma = \set{\text{cavity photon, CH\textsubscript{2}, CF\textsubscript{2}}}$. The $\sigma$-component in a polariton state $\ket{\Psi}$ is computed as
$
P^\sigma = \braket{\Psi|\mc{P}_\sigma |\Psi}
$.  As shown in \fig{fig:xas},
without cavity ($g=0$), all excitations are either purely CH\textsubscript{2} (yellow) or CF\textsubscript{2} (purple) type.
As the coupling is turned on, the two \around \SI{290}{\electronvolt} excitation contain mixed CF\textsubscript{2} and photon (brown) character, reflecting a hybridization of the CF\textsubscript{2} $\rightarrow \pi^*$ and the cavity photon, resembling the polariton states in a Jaynes-Cummings model. As $g$  increases, the polariton states further mix with CH\textsubscript{2}-excitations, leading to {delocalized} core-excitations from both CH\textsubscript{2} and CF\textsubscript{2}. The delocalization  can be clearly observed in the decomposition of the polariton states \around \eV{290}.
These delocalized excitations involving both C atoms arise from an effective coupling between their core excitations induced by exchanging cavity photons even when the cavity is in the vacuum state.
%
%
When the cavity frequency is detuned far from any resonance in the bare XANES $\omega_\text{c} = \eV{288}$ (bottom row of \fig{fig:xas}), {no} substantial changes in the spectrum {are observed} at $g =$ 2.45 eV/D. Nevertheless, as $g$ {gets stronger}, we observe similar delocalized core-excitations involving both CH\textsubscript{2} and CF\textsubscript{2} at e.g. \eV{290}.

\begin{widetext}

\begin{figure}[ht]
	\includegraphics[width=0.8\textwidth]{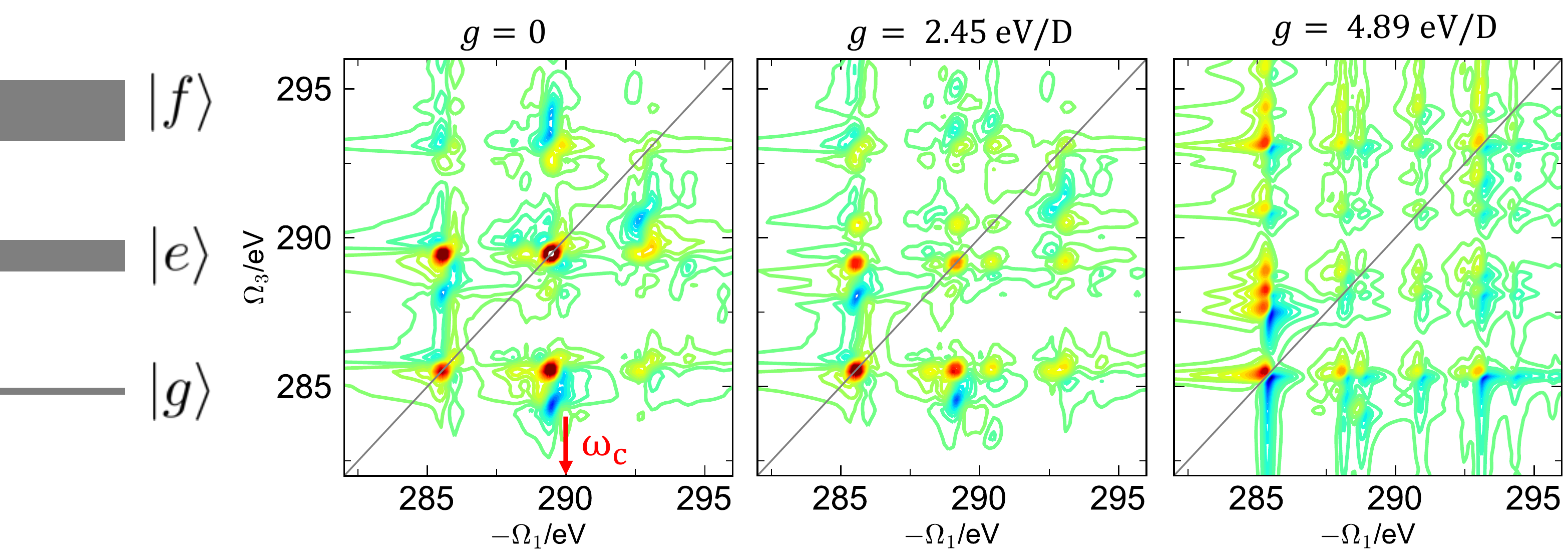}
	\caption{ Level scheme and the 2D photon echo spectra {$S_\text{PE}(\Omega_3, \Omega_1; T_2 = 0)$} for  1,1-difluoroethylene in an X-ray cavity with $\omega_\text{c} = \eV{290}$ for different coupling strengths as indicated. {We use attosecond pulses with central frequency \eV{290} and  \eV{20} bandwidth.}}
	\label{fig:pe}
\end{figure}

\end{widetext}

For nonlinear X-ray signals,  we focus on the single-molecule  $N=1$ strong coupling case. {Single-molecule strong coupling requires a substantial field enhancement, and its realization may benefit from an  ensemble of auxiliary emitters \cite{Schutz2020a}.}  {The signals for large $N$ can depend on many collective dark states that are neglected here.} Doubly core-excited dark states $\ket{e_\alpha^\mol e_\beta^{(m)}}$ {in different molecules} {also need to be taken into account}. Such states do not show up in {bare} nonlinear spectroscopy due to destructive interference \cite{Muthukrishnan2004, Richter2011}. The cavity mode mediates an effective coupling even for otherwise non-interacting molecules, and the two-core-exciton states from different molecules  will influence the bipolariton manifold.   {Below $g, e, f$ label the ground, single-polariton, two-polariton states, respectively, {see \fig{fig:pe} for the level scheme.}}

We have computed time-domain heterodyne-detected 2D X-ray four-wave mixing signals of core-polaritons. These allow us to track the time-evolution of the polariton states and reveal correlations between transitions.  The total electric field is decomposed into three pulses
  $ E(t) = E_3(t) + E_2(t + T_2) + E_1(t + T_1 + T_2) + \text{c.c.}
  $
where $T_j$ is the time-delay between the $j$-th and $j+1$-th pulse.
Labeling the wave vectors of the incoming pulses as $\bf k_j$,  we first discuss photon echo (PE) signal at $- \bf k_1 + \bf k_2 + \bf k_3$.


The 2D PE spectra are  sketched by the Liouville space pathways represented by  Feynman diagrams \cite{Mukamel1995}, depicted in Fig. S1.
 The 2D correlation spectra are obtained by Fourier transforming the delays $T_1$ (coherence time) and $T_3 \equiv t$ (detection time) {in the polarization}
 $
 S_\text{PE}(\Omega_3, \Omega_1; T_2) = \int_0^\infty \dif T_1 \int_0^\infty \dif T_3\braket{P_\text{PE}(T_3, T_2, T_1)} e^{i \Omega_3 T_3 + i \Omega_1 T_1}
 $.

The 2D PE signals are displayed in \fig{fig:pe}. {There are three contributions to the spectra:} stimulated emission (SE) and gound-state bleaching (GSB) (the first two diagrams in Fig. S1), and the excited state absorption (ESA, the last diagram in Fig. S1). {The four XANES features discussed earlier give rise to four traces along $\Omega_1$ (i.e., CH\textsubscript{2} excitations at \SI{285.6}{\electronvolt}, \SI{288.4}{\electronvolt} and \SI{293.0}{\electronvolt} and CF\textsubscript{2} excitations at \SI{289.5}{\electronvolt}) with a characteristic cross peak pattern, that reflects the correlation between various transitions. The cross peaks result from the fact that they share a common ground state, and that  the core excitations are both anharmonic $\omega_{fe} \ne \omega_{eg}$ and coupled $\omega_{fg} \ne  \omega_{eg} + \omega_{e'g}$.
	ESA signals related to double core-excitations from the same carbon atom do not cancel the respective GSB and SE signals, consequently, cross-peaks appear symmetrically below and above the diagonal.
	 Transitions involving CH\textsubscript{2} and CF\textsubscript{2} cores are quartically coupled due to  spatial vicinity of the two carbons, i.e., excitations of CH\textsubscript{2} core depends on {the occupation number in CF\textsubscript{2}}. The associated ESA exhibit a $\sim$1.5 eV red-shift (\SI{289.6}{\electronvolt}/\SI{284.0}{\electronvolt} and \SI{285.6}{\electronvolt}/\SI{288.0}{\electronvolt}) or a blue-shift (\SI{289.6}{\electronvolt}/\SI{294.5}{\electronvolt} and \SI{293.0}{\electronvolt}/\SI{291.0}{\electronvolt}) with respect to the corresponding off-diagonal GSB which makes the ESA appear in the 2D spectra.}
 At $g = 2.45 $ eV/D, the   polariton splitting is reflected in the additional cross peaks between the polariton states and bare molecular states. Similar features {are seen} in the stronger coupling case shown in \fig{fig:pe} where additional hybrid polariton states containing excitations from both carbon atoms are created.

We now turn to the double quantum coherence (DQC) signal at $\bf k_1 + \bf k_2 - \bf k_3$ \cite{Richter2010, Schweigert2008, Kim2009, Dai2012, Abramavicius2012}, {represented by the diagrams shown in Fig. S2}.
 The correlations between single- and two-polaritons can be obtained either by Fourier transform of the time-delays $T_3$ and $T_2$  at a fixed  $T_1$,
$
S_\text{DQC}(\Omega_3, \Omega_2; T_1)$
or
by Fourier transform of the time-delays $T_1$ and $T_2$ at a fixed $T_3$
$
S_\text{DQC}(\Omega_2, \Omega_1; T_3)$.
In DQC, the polariton system is in the coherence $\ket{e}\bra{g}$ during $T_1$, and is then promoted to $\ket{f}\bra{g}$ during $T_2$. The system can be either $\coh{f}{e}$ or $\coh{e}{g}$ for the detection time $T_3$. The peaks in $S_\text{DQC}(\Omega_2, \Omega_1; T_3)$ reveal correlation between $\omega_{eg}$ and $\omega_{fg}$.  For a harmonic system where $\omega_{fg} = \omega_{eg}$ and for uncorrelated transitions where $\omega_{fg} = \omega_{eg} + \omega_{e'g}$, the DQC signal vanishes as the two contributions to DQC interfere destructively. This makes DQC suitable for resolving anharmonicities and correlated transitions.

The DQC  $S_\text{DQC}(\Omega_2, \Omega_1; T_3)$ are shown in \fig{fig:DQC} for varying cavity coupling strengths. The vertical axis shows the doubly core-excited states $\ket{f}$ that can be reached from $\ket{g}$ through an excited state $\ket{e}$. {Prominent contributions at $\Omega_1/\Omega_2 = \SI{285.6}{\electronvolt}/\SI{573.9}{\electronvolt}$ and $ \SI{289.5}{\electronvolt}/\SI{573.9}{\electronvolt}$ arise due to the coupling of CH\textsubscript{2}$\rightarrow \pi^*$ (\SI{285.6}{\electronvolt}) and CF\textsubscript{2} $\rightarrow \pi^*$ (\SI{289.5}{\electronvolt}) transitions to the CH\textsubscript{2},CF\textsubscript{2} $\Rightarrow \pi^*$ transition
	\footnote{Here $\Rightarrow$ indicates a double excitation.}. Similarly, peaks at $\SI{289.5}{\electronvolt}/\SI{584.1}{\electronvolt}$ and $\SI{293.0}{\electronvolt}/\SI{584.1}{\electronvolt}$ arise due to the coupling of CF\textsubscript{2} $\rightarrow \pi^*$ (\SI{289.5}{\electronvolt}) and CH\textsubscript{2}$\rightarrow $ Ry (\SI{293.0}{\electronvolt}) transitions to the CH\textsubscript{2}, CF\textsubscript{2} $\Rightarrow \pi^*$, Ry transition. } In the strong coupling regime, the polariton states manifest as a doublet around $\Omega_1 = \eV{290}$. Additional peaks are clearly observed between these single-polariton states and the $f$-manifold. Core-polaritons also modulate the doubly core-excited states {by} mixing {them with the} two-cavity-photon state and single-core-excitation single-cavity-photon state. For example, a noticeable red-shift can be  observed for the CH\textsubscript{2},CF\textsubscript{2} $\Rightarrow \pi^*$  transition from the slices of the DQC at $\Omega_1 = \eV{285.5}$.  
%

The correlations between $\omega_{fe}$ and $\omega_{fg}$  are revealed in the DQC signal $S_\text{DQC}(\Omega_3, \Omega_2; T_1)$ displayed in \fig{fig:DQC} (bottom row). 
{States from the doubly excited manifold  coupled to the singly excited manifold are characterized through a set of four peaks along $\Omega_3$ for a given $\Omega_2$ value \cite{Nenov2014}.  For example, the quartet of peaks along the $\Omega_2 = \SI{573.9}{\electronvolt}$ are associated with two peaks at \SI{285.6}{\electronvolt} and \SI{289.5}{\electronvolt} coinciding with the CH\textsubscript{2}$\rightarrow \pi^*$ and CF\textsubscript{2} $\rightarrow \pi^*$ transitions 
	and two red-shifted peaks at \SI{284.4}{\electronvolt} and \SI{288.3}{\electronvolt} corresponding to the CH\textsubscript{2}$\rightarrow \pi^*$  with  CF\textsubscript{2} excited and CF\textsubscript{2} $\rightarrow \pi^*$ with  CH\textsubscript{2}-excited.
	\footnote{Here, the brackets indicate that the core-excitation occurs in the presence of a core-excited carbon.}
	 The \SI{1.2}{\electronvolt} splitting between each pair of corresponds to the value of the quartic coupling between both transitions.} {Under strong coupling,} core-polariton doublets can be observed along $\Omega_3$ due to the $\omega_{fe}$ resonances.  The splitting does not directly correspond to the polariton resonances because both $e$ and $f$ manifolds are modified by the cavity mode.

{Similar information about the correlations of single and double excitations as provided by DQC can be extracted from the two-photon absorption signal, discussed in Sec. S4. This signal does not require coherent X-ray pulses and is thus easier to implement experimentally.}

In summary, we have demonstrated how molecular core-excitations can be manipulated by coupling to the vacuum field in an X-ray cavity.
Localized excitations from the two carbon atoms in 1,1-difluoroethylene are coherently coupled by the exchange of an X-ray cavity photon creating hybrid delocalized excitations.
We identified the spectroscopic signatures of core-polaritons in XANES,  two-photon absorption, and multidimensional X-ray spectroscopic signals.
XANES directly probes the hybrid core-polariton states with the polariton effects manifested as mode splitting, redistribution of oscillator strength, and line shifts, depending on the cavity frequency and coupling strength.
Correlation between polaritonic excitations are revealed by the PE, and
information about the two-polariton manifold can be readily extracted from the DQC and two-photon absorption signals.
This study shows how to  manipulate core-excitations in molecules by strong coupling to a cavity in the X-ray regime. { Many interesting phenomena discovered for exciton-polaritons in the optical regime such as long-range transport \cite{Rozenman2018}, modified chemical reaction rates \cite{Hutchison2012}, enhanced nonlinearity \cite{Chervy2016} suggest analogous extensions for core-polaritons in the X-ray regime. {Relaxation dynamics of core-polaritons is also expected to differ significantly from the bare core-excitations.} {Collective effects found in M\"{o}ssbauer resonance in iron including electromagnetically induced transparency and Lamb shift \cite{Rohlsberger2012, Rohlsberger2010} may show up in molecules as well.}



\begin{figure}[ht]
	\includegraphics[width=0.48\textwidth]{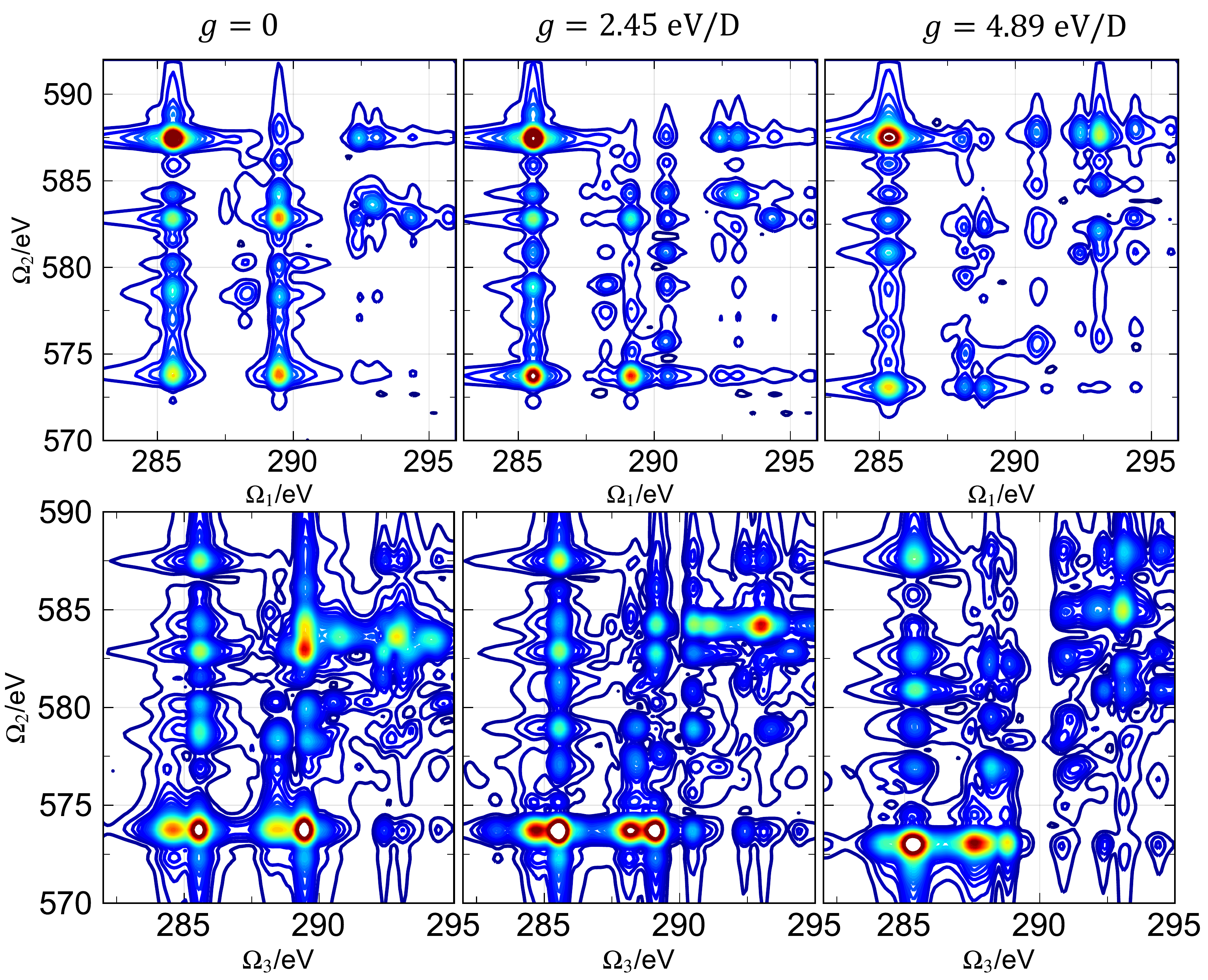}
	\caption{2D double quantum coherence spectra $\abs{S_\text{DQC}(\Omega_2, \Omega_1; T_3)}$ (upper row) and $\abs{S_\text{DQC}(\Omega_3, \Omega_2; T_1)}$ (lower) in an X-ray cavity with $\omega_\text{c} = \eV{290}$ at different coupling strengths $g$ as indicated.  {A small time delay $ \num{d-5}$ as is used for both $T_3$ and $T_1$ to avoid  cancellation of the two DQC diagrams.}}
	\label{fig:DQC}
\end{figure}

\begin{acknowledgments}
	We thank Dr. Stefano M. Cavaletto  for helpful discussions \added{and Dr. Ralf R\"{o}hlsberger for valuable feedback}.
M.G., A.N., F.S., and S.M. acknowledge support from the Chemical Sciences, Geosciences, and Bio-sciences division, Office of Basic Energy Sciences, Office of Science, U.S., Department of Energy, through Award No. DE-SC0019484. B.G. acknowledges the
support from the National Science Foundation Grant CHE-1953045.
\end{acknowledgments}

\end{document}